\documentclass[a4paper,10pt]{article}
\usepackage[utf8]{inputenc}

\title{Boolean Logic Gates From A Single Memristor Via Low-Level Sequential Logic}
\author{Ella Gale, Ben de Lacy Costello and Andrew Adamatzky}

\usepackage{amssymb}
\usepackage{graphicx}

\begin{document}

\maketitle

\begin{abstract}
By using the memristor's memory to both store a bit and perform an operation with a second input bit, simple Boolean logic gates have been built with a single memristor. The operation makes use of the interaction of current spikes (occasionally called current transients) found in both memristors and other devices. The sequential time-based logic methodology allows two logical input bits to be used on a one-port by sending the bits separated in time. The resulting logic gate is faster than one relying on memristor's state switching, low power and requires only one memristor. We experimentally demonstrate working OR and XOR gates made with a single flexible Titanium dioxide sol-gel memristor.
\end{abstract}

\section{Introduction}

The memristor is the recently-discovered~\cite{Strukov} fourth fundamental element, joining the set of the resistor, inductor and capacitor. It was predicted to exist based on an expectation of symmetry in electromagnetic phenomena when applied to circuit theory~\cite{Chua1971}, specifically in that it would be passive two-terminal device that would relate the two as-then-unrelated circuit measurables: charge, $q$, and magnetic flux, $\varphi$~\footnote{The other measurables being current and voltage, and the other five relationships being the definitions of current and voltage and the constitutive relations of the other three fundamental circuit elements}. From knowledge about its electronic properties, Chua predicted that it would be a non-linear version of a resistor that possesses a memory, hence the name memristor, a contraction of memory-resistor.

Whilst Chua's theoretical contributions were not known to the wider chemical and physics communities, devices highly similar in constitution and operation to Strukov's memristor~\cite{Strukov} were created and dubbed ReRAM, for Resistive Random Access Memory, after the use their inventors intended for them. What exactly constitutes a memristor or ReRAM device is a matter of debate, although it has been suggested that they may be the same thing~\cite{Chua2011}. Both memristors and ReRAM have suggested uses as computer memory and both are believed to possess the same physical interactions and thus, in this paper, we shall deal with both under the name of memristors, where it is understood that a large part of the results presented here should be tested on ReRAM devices and are expected to work in the same way.

Both memristors~\cite{Strukov} and ReRAM~\cite{Waser} have been suggested as possible low-power next-generation computer memory technology, however the field of ReRAM has been around for 20 years and has not yet produced a commercial product and Hewlett-Packard (the company that discovered the Strukov memristor) has been delaying their computer memory offering based on their memristor. 

Chua's theoretical model of the memristor has been used to model neuronal synapses (see for example~\cite{110,239,DavidJ1}) and to update the Hodgkin-Huxley model of neuronal membranes and axonal transport~\cite{247,248}. It has been shown~\cite{ICNAAM} that the experimental memristor spikes in a similar manner to those seen in axonal transport, where it is understood that neurons demonstrate a voltage spike in response to the current influx, and the spikes shown in~\cite{ICNAAM} are current spikes in response to the voltage change. These current spikes have been seen in other memristor systems to ours and are generally ignored or dismissed as current transients. A view which will be tested with our devices in a forthcoming paper. Regardless of how these spikes arise, they are impossible (as far as we know and the literature states) to remove and thus it is our opinion that future uses of memristor technology will have to involve these spikes. Based on the relation to the brain's operation, we consider that 
memristor networks will be useful for neuromorphic computing, however in this paper we will demonstrate how a single memristor can be used as a Boolean logic gate by making use of the physical property of these current spikes, which can be done if we take an unconventional approach to logic assignation.

This is not the first paper on how to make logic gates with memristors. Strukov et al~\cite{242} resorted to using implication logic to design logic gates which required two memrsitors (IMP-FALSE logic is Turing complete, but somewhat unfamiliar to computer scientists). The most notable Boolean logic gates were simulated by Pershin and di Ventra~\cite{PAndV} and required a memcapacitor, three or four memristive systems and a resistor. Before the gate was sent the two bits of data, a set of initialization pulses were required to be sent to put the gate into the correct state to give the correct answer. This system, however, is not true Boolean logic because these initialization pulses were different dependent on what the logic to follow would be. Thus the gate can not be considered to be operating only on the two bits of input data and is not a simple Boolean logic gate (it is a Turing machine doing a computation on several bits of data (Boolean input pulses and initialization pulses) which is capable of modeling a Boolean logic gate). Note also that this scheme was tested with memristor emulators, not real devices. There have been other more complex designs for memristor based Boolean logic gates, the simplest of which requires 11 circuit elements~\cite{Pino}. In this paper, we will demonstrate how to perform Boolean logic with a single memristor.

Although the memristor is credited with being the first computational device to combine memory and processing functionality in one, with the suggestion of an entirely new type of computer, remarkably there have been relatively few papers on how this new computer might work: most people have chosen to focus on stateful memory applications~\cite{242}.

We will now demonstrate the physical properties of our memristors~\cite{MemMethod} and validate their reproducibility (section~\ref{sec:repro}), explain the concept of how these spikes can be used to perform Boolean logic (section~\ref{sec:logic}) and, as an example, demonstrate experimentally that a single memristor can act an OR (section~\ref{sec:OR}) or and XOR (section~\ref{sec:XOR}) gate.

\section{Methodology}

The memristors are flexible TiO$_2$ sol-gel memristors with aluminium electrodes and were made as in~\cite{MemMethod} with the sol-gel created as in~\cite{MatChemSci} (the memristor chosen was a curved-type memristor (see~\cite{MemMethod})). All tests on the memristor were performed with a Keithley 2400 Sourcemeter and data was recorded and analysed using MatLab. Each timestep was 0.02s. The voltages used and voltage waveforms varied as are discussed below.

\section{Physical Properties of the Memristor~\label{sec:repro}}

When there is a change in voltage, $\Delta V$, across a memristor the device exhibits a current spike, the physical cause of which is discussed at length in~\cite{ICNAAM}. This spike is highly reproducible and repeatable and is related to the size of the voltage change ($\Delta V$)~\cite{ICNAAM}. The spike's size (as measured by the first measurement after the Keithley's changed voltage) is highly reproducible, the current then relaxes to a stable long-term value (this value is predictable and reproducible), and it takes approximately 2-3 seconds to get to this value. 

This slow relaxation is thought to be the d.c. response of the memristor~\cite{ICNAAM} and if a second voltage change happens within this time frame, its resulting current spike is different to that expected from the $\Delta V$ alone. The size and direction of this current spike depends on the direction of $\Delta V$, the magnitude of $\Delta V$ and the short-term memory of the memristor. 
 
\begin{figure}[htbp!]
 \centering
 \includegraphics[bb=0 0 576 432,scale=0.5,keepaspectratio=true]{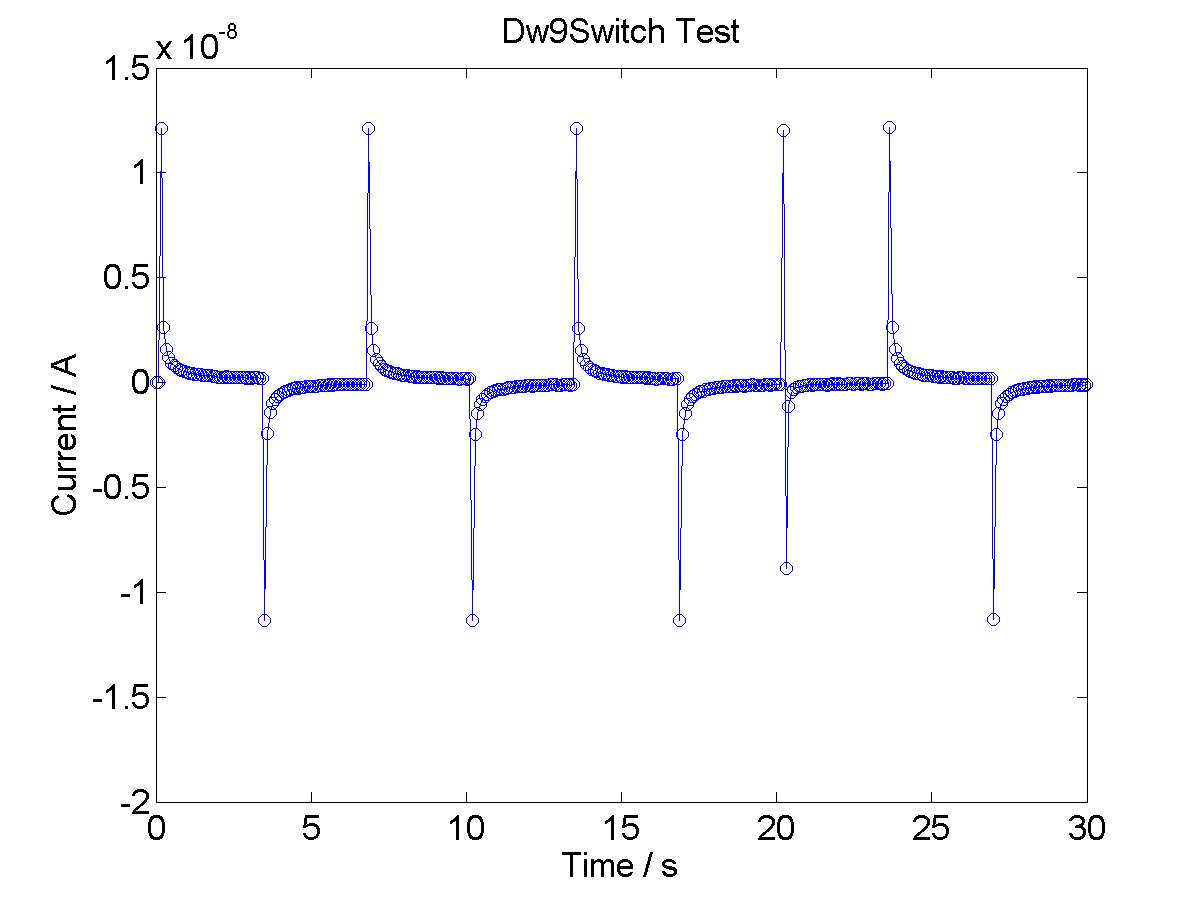}
 \caption{The effect of adding spikes close in time. The response spikes are the negative current spikes. When a positive spike it included but not allowed to relax the corresponding negative spike is smaller.}
 \label{fig:Test1Cu}
\end{figure}

\begin{figure}[htbp!]
 \centering
 \includegraphics[bb=0 0 576 432,scale=0.5,keepaspectratio=true]{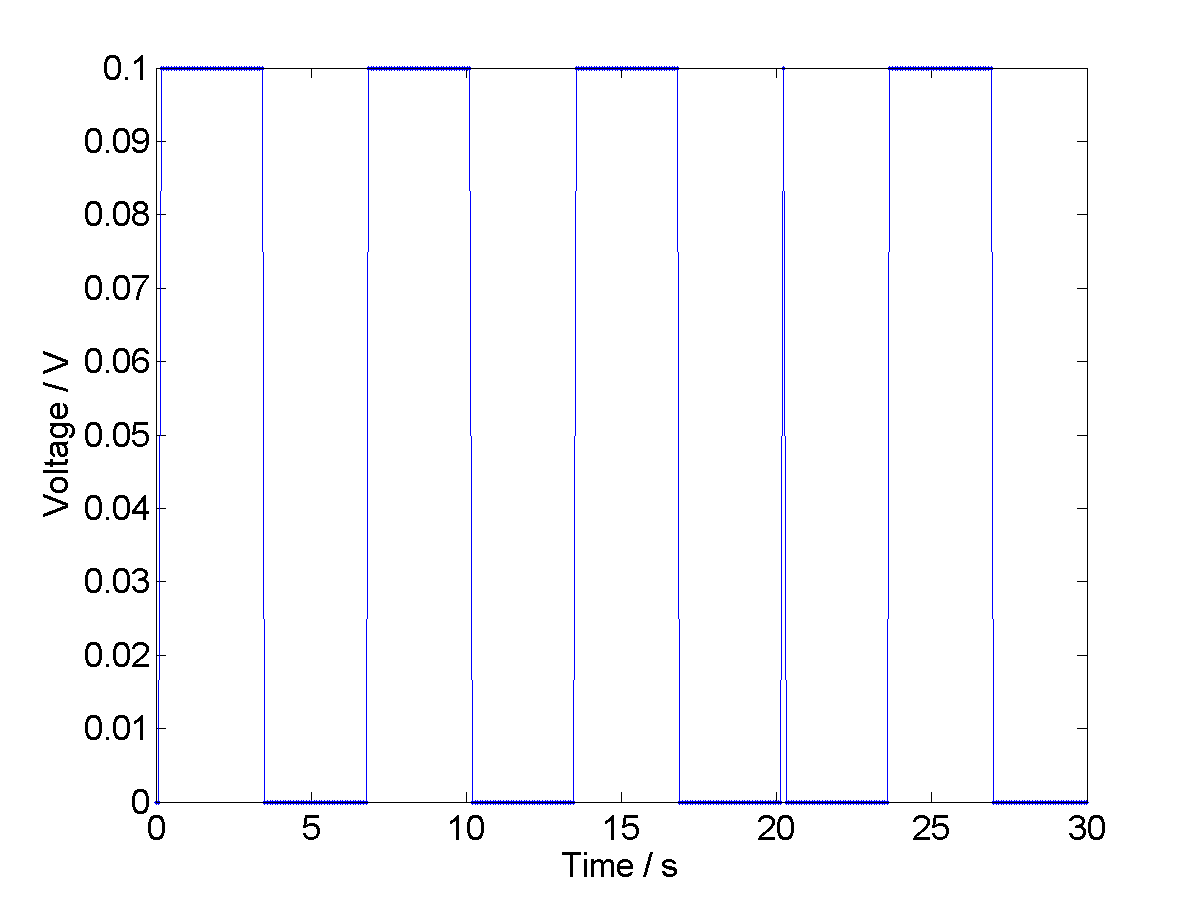}
 \caption{The input voltage for Test 1.}
 \label{fig:Test1V}
\end{figure}

As an example, consider a memristor pulsed with a positive 1V voltage square wave as in figure~\ref{fig:Test1V} (where the pulses are repeated to demonstrate the repeatability) with a timestep of $\approx 0.02s$. The current response is shown in figure~\ref{fig:Test1Cu} and we can see there is a positive current spike associated with the $+\Delta V$ and a perhaps less obvious negative current spike associated with the $-\Delta V$ transition from $+1V \rightarrow 0V$. At approximately 20s, we shortened the square wave to a single time step, and the memory of the system has caused the response spike (responding to the $-\Delta V$ to be smaller (and as it is smaller, it suggests that there is some physical property of the device which has not adjusted to its $+V$ value. See~\cite{Gale} for an discussion on why this physical property is predicted to be the oxygen vacancies in the TiO$_2$.) Thus the response is subtractive in current and additive in resistance state.

\begin{figure}[htbp!]
 \centering
 \includegraphics[bb=0 0 576 432,scale=0.5,keepaspectratio=true]{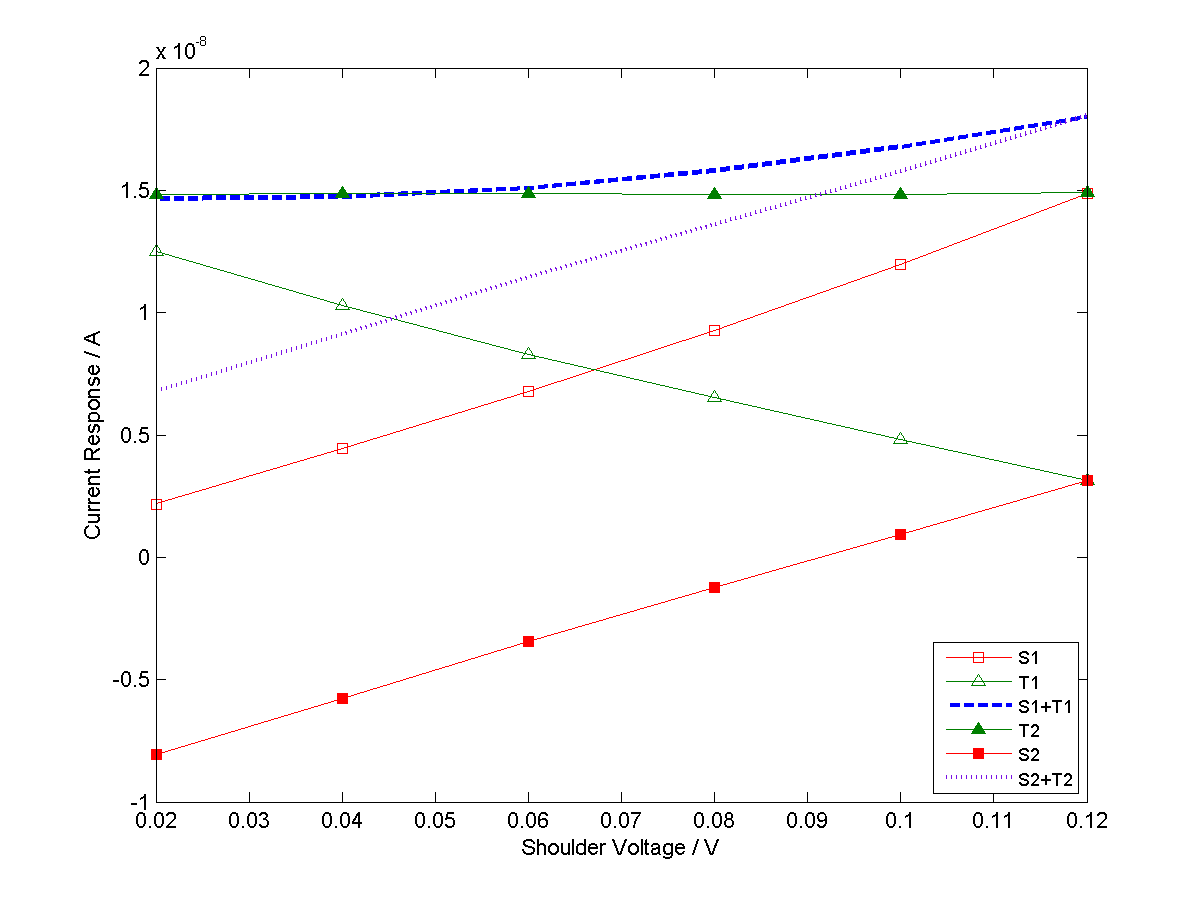}
 \caption{The effects of the order the spikes are sent in to show that spike addition is non-commutative. S1=$\Delta i_{0 \rightarrow A}(t)$, T1=$i_{ A \rightarrow B}(t+1)$, T2=$\Delta i_{0 \rightarrow B}(t)$ and S2=$\Delta i_{B \rightarrow A}(t+1)$. S1 and T1 refer to the shoulder (S1) and peak (T1) currents resulting switching from $0V\rightarrow V_A \rightarrow V_B$. T2 and S2 refer to the peak (T2) and shoulder (S2) of switching from $0V\rightarrow V_B \rightarrow V_A$. In both cases $V_B > V_A$.}
 \label{fig:test10}
\end{figure}

To try and understand the subtleties of this apparent `addition', consider the following system: two voltages are sent to the memristor, one after the other separated by one timestep (i.e. before the memristor has equilibrated), where $V_B > V_A$ and $V_B = 0.12V$, and figure~\ref{fig:test10} shows the size of the two resulting spikes as a function of increasing $V_A$. We look at two situations: 1, $V_A(t) \rightarrow V_B(t+1)$; 2, $V_B(t) \rightarrow V_A(t+1)$. These two situations are drastically different if we look at the transitions, $\Delta V$, as situation 2 has a negative $\Delta V_{B \rightarrow A}$, all the other transitions are positive. Situation 1 shows that if the smaller voltage is sent first ($V_A \rightarrow V_B$), the current of the first transition $\Delta i_{0 \rightarrow A}$ increases with the size of $V_A$, and the second transition $\Delta i_{A \rightarrow B}$ decreases with the size of $\Delta V_A$, due to the decrease in the effective $\Delta V_{A \rightarrow B}$. However, the sum of 
these two effects is non-linear, so that the total current transferred (approximated as the sum of the spikes here, but actually the area under the two current transients) is not the same as that shown for situation 2 (until $V_B=V_A$). This shows that more current is being transferred and demonstrates that the spikes are dependent on $\Delta V$. Furthermore, it makes it clear that $\Delta i_{0 \rightarrow A} + \Delta i_{A \rightarrow B} \neq \Delta i_{0 \rightarrow B} + \Delta i_{B \rightarrow A}$, (except in the trivial case where $V_B = V_A$) and that spike based `addition' is non-commutative and therefore the order in which the spikes are sent is relevent.  

\section{Boolean Logic Using Current Spikes in Memristors~\label{sec:logic}}

We can do Boolean logic with the spike interactions by sending the second bit of information one timestep (0.02s) after the first. We take the input as the current spikes from the voltage level. The output is the response current as measured after the 2$^{\mathrm{nd}}$ bit of information. After a logic operation the device is zeroed by being taken to 0V for approximately 4s, and this removes the memristor's memory. 

We have some freedom in how we assign the `1' and `0' states to device properties and these give different logic. The following examples will demonstrate some approaches and build an OR gate or an XOR gate.

\subsection{OR Gate~\label{sec:OR}}

\begin{table}[htb]
\caption{OR Truth Table (inclusive OR)}
 \begin{tabular}{|l|l|c|}
  \hline
  Input 1	& Input 2	& Output	\\
  \hline
  0		& 0		& 0\\
  \hline
  0		& 1		& 1\\
  \hline  
  1		& 0		& 1\\
  \hline
  1		& 1		& 1\\
  \hline
 \end{tabular}
\label{tab:OR}
\end{table}

The truth table for an OR gate is given in table~\ref{tab:OR}, essentially, the output should be `1' if either of the inputs was `1'. We take the `0' output as being below a threashold current and the `1' output as being above a threashold. The threashold is set to $>$18nA with the `0' input being set of 0.01V and the `1' as 0.2V~\footnote{Using `0' as 0V was also tested, it works and is lower power but was not chosen as an example as it is a trivial case.}, which gives the voltages below:
 
\begin{itemize}
\item 0, 0 = 0.01V, 0.01V
\item 0, 1 = 0.01V, 0.2V
\item 1, 0 = 0.2V, 0.01V
\item 1, 1 = 0.2V, 0.2V.
\end{itemize}

\begin{figure}[htbp!]
 \centering
 \includegraphics[bb=0 0 576 432,scale=0.5,keepaspectratio=true]{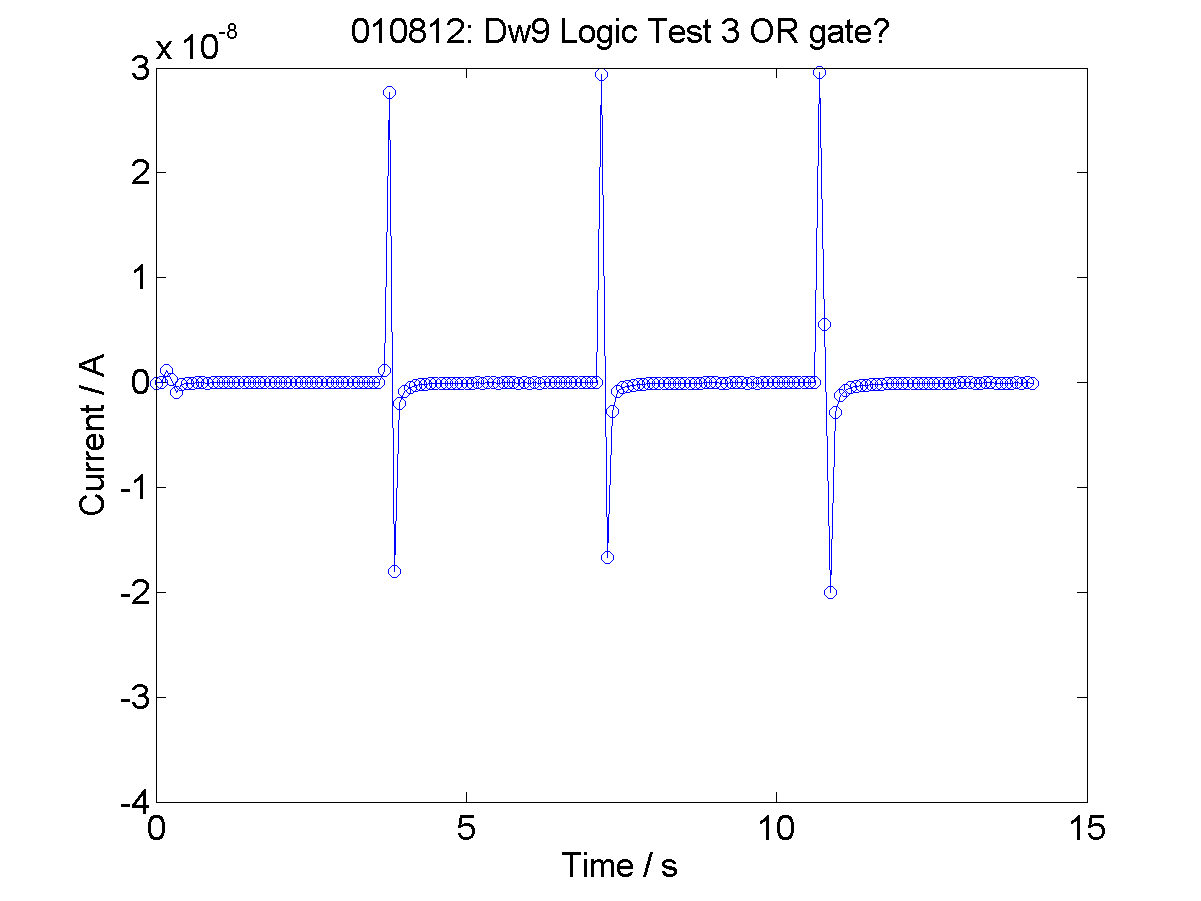}
 \caption{OR Gate. Using `1' equal to a current spike caused by a voltage change to 0.2V and `0' equal to a current spike caused by a voltage change to 0.01V we can make a serial OR gate (where logical 1 is considered to be a current which is more than 5nA). At 0.04s `0, 0' was input, giving peaks below the threashold i.e. `0' as an output. The three large peaks are `1' outputs resulting from `0,1',`1,0' and `1,1' inputs.}
 \label{fig:LogicTest3}
\end{figure}

Figure~\ref{fig:LogicTest3} shows the current data from the voltage inputs above. It can be seen that when a `1' is input, there is a large spike output. To read the logical state of the device, one merely takes the current value as the second bit is read in. 

\subsection{A Logical System to Create an XOR Gate~\label{sec:XOR}}

The XOR truth table is shown in table~\ref{tab:XOR}. If we take logical `1' to be the current resulting from a positive voltage and a logical `0' to be the current resulting from a negative voltage, then, the response is the current when the 2$^{\mathrm{nd}}$ bit is input (not after, although it could be designed that way but it is slower). We get a high absolute value of current if and only if the two inputs are of different signs, i.e. we have $1\:0$ or $0\:1$ which gives us an exclusive OR operation. For this logical system, we used the same voltage level and allowed a change in sign to indicate logical zero or logical one:
\begin{itemize}
\item 0, 0 = -0.1V, -0.1V
\item 0, 1 = -0.1V, +0.1V
\item 1, 0 = +0.1V, -0.1V
\item 1, 1 = +0.1V, +0.1V.
\end{itemize}

\begin{table}
\caption{XOR Truth Table (exclusive OR)}
 \begin{tabular}{|l|l|c|}
  \hline
  Input 1	& Input 2	& Output	\\
  \hline
  0		& 0		& 0\\
  \hline
  0		& 1		& 1\\
  \hline  
  1		& 0		& 1\\
  \hline
  1		& 1		& 0\\
  \hline
 \end{tabular}
\label{tab:XOR}
\end{table}

\begin{figure}[htbp!]
 \centering
 \includegraphics[bb=0 0 576 432,scale=0.5,keepaspectratio=true]{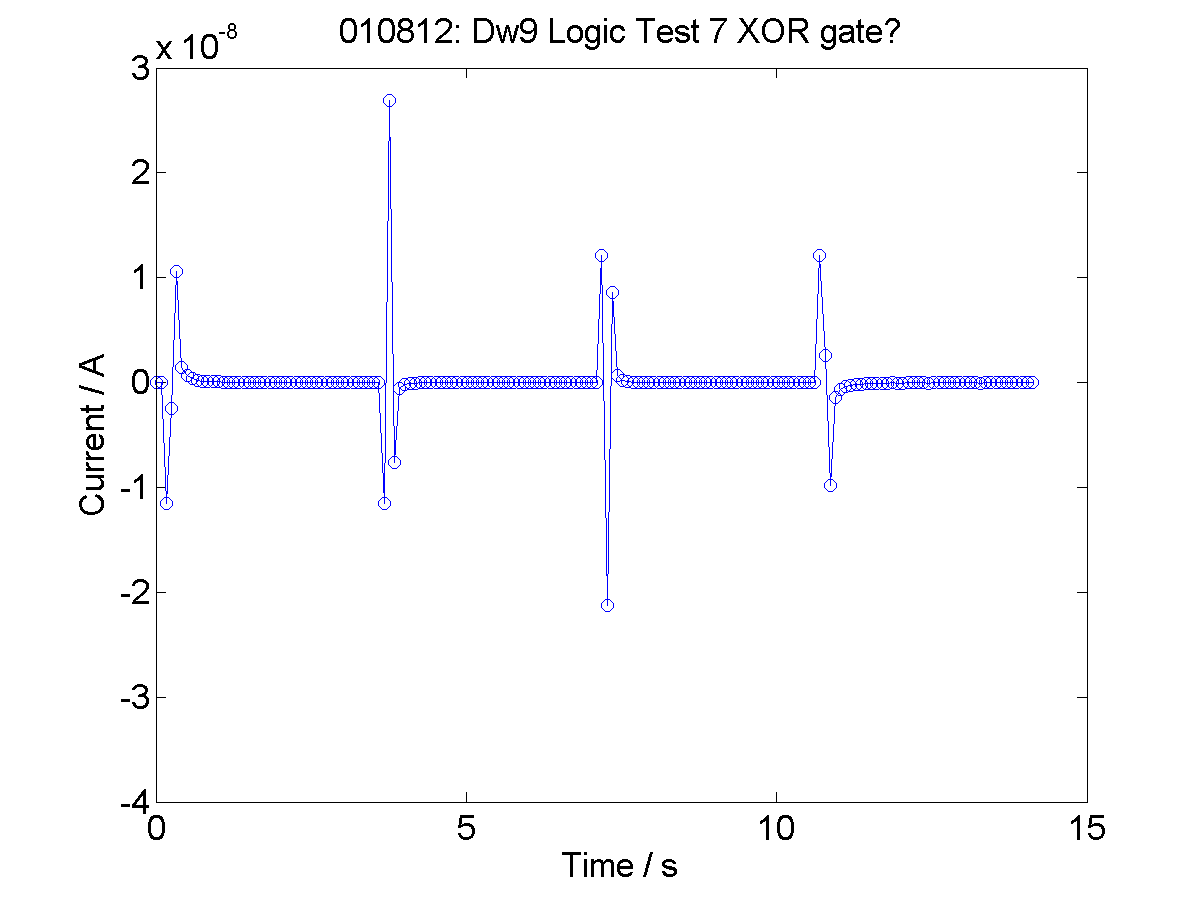}
 \caption{XOR gate, where a current response over $\pm 1.25\times10^{-8}$A is taken as one, as current response under that threashold is taken as zero.}
 \label{fig:Test7Current}
\end{figure}

\begin{figure}[htbp!]
 \centering
 \includegraphics[bb=0 0 576 432,scale=0.5,keepaspectratio=true]{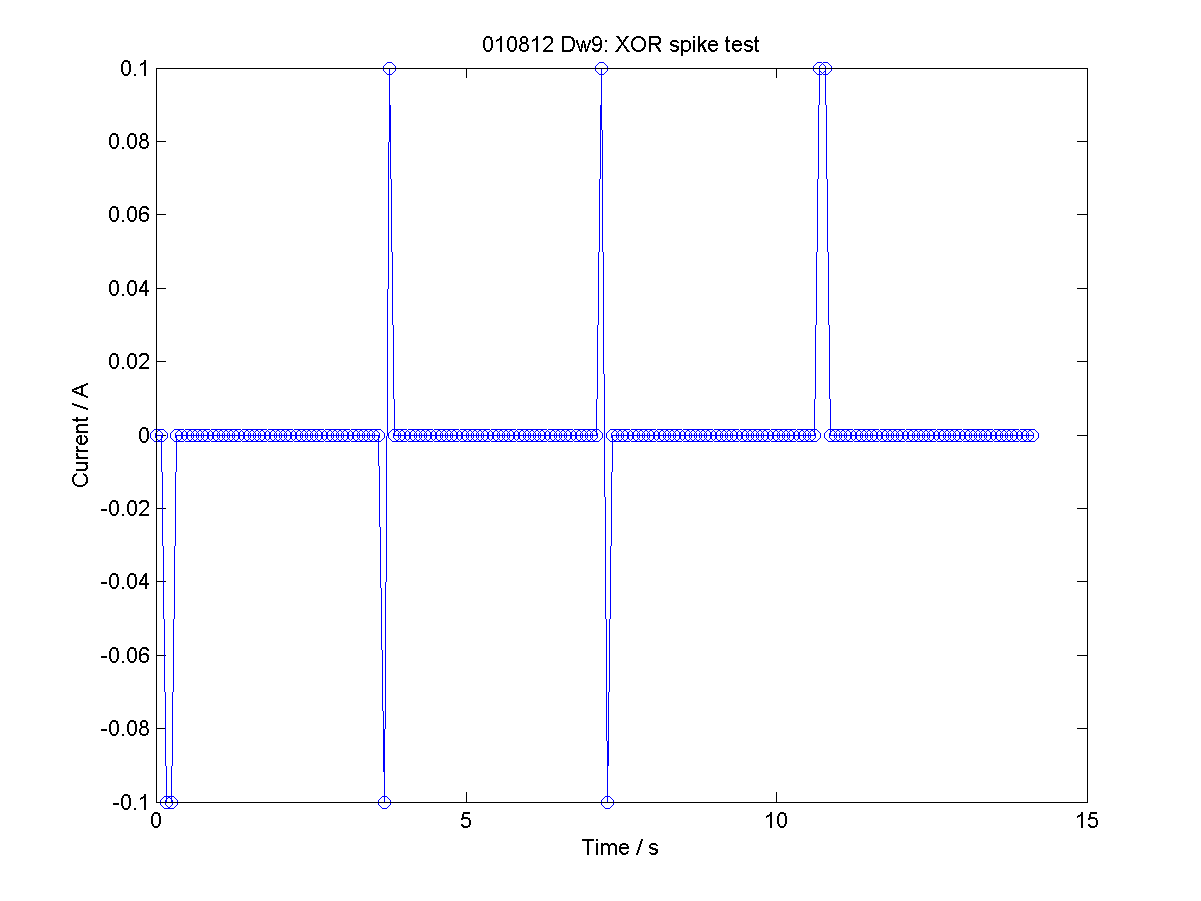}
 \caption{The programming voltage for the XOR gate.}
 \label{fig:VoltageXOR}
\end{figure}

As an example, the input voltage is shown in figure~\ref{fig:VoltageXOR} and the current output is shown in figure~\ref{fig:Test7Current}.

\begin{figure}[htbp]
 \centering
 \includegraphics[scale=0.5,keepaspectratio=true]{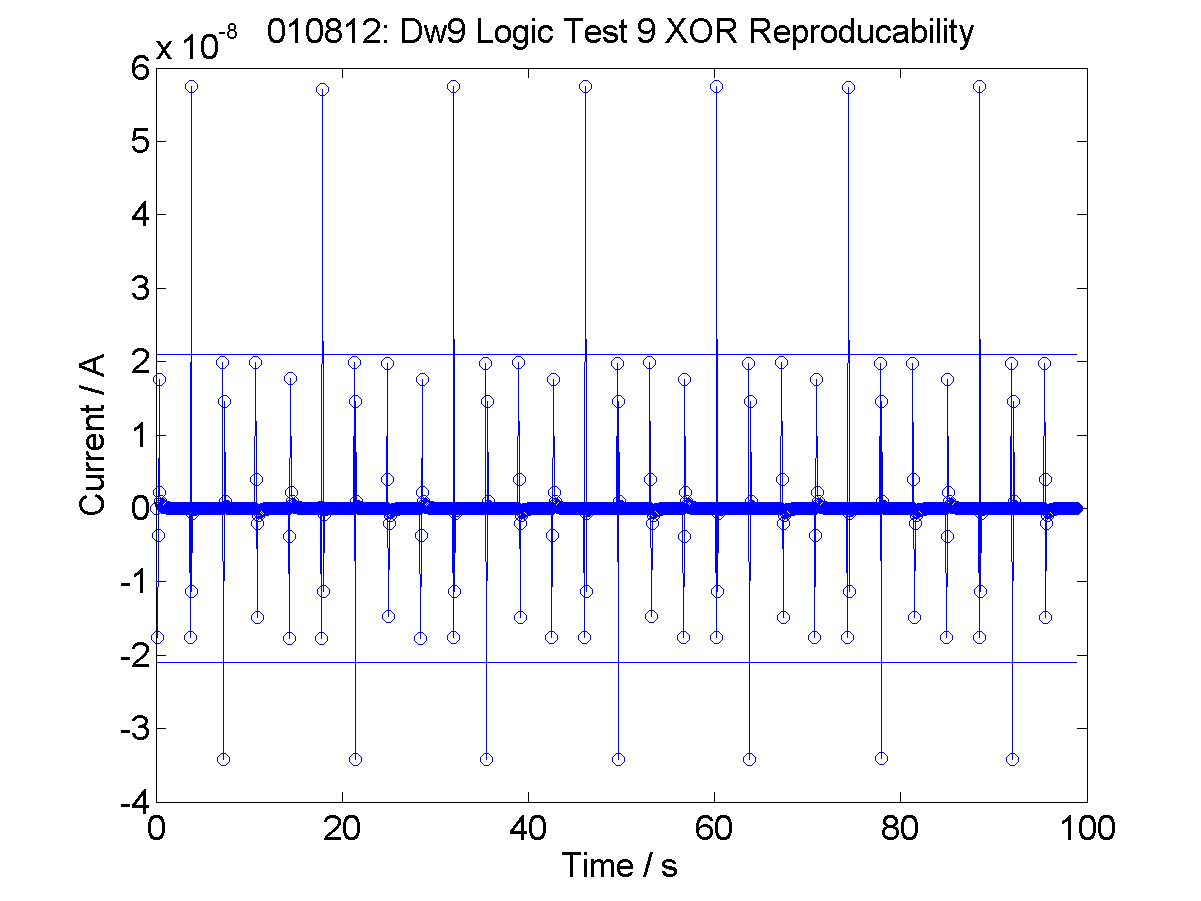}
 \caption{Reproducabiltiy test of XOR function. Here the XOR truth table is run 7 times. The threshold between `1' and `0' is marked as shown.}
\label{fig:XORReproTest}
\end{figure}

With a pause between operations to allow the memristor to lose its memory, the XOR operation is reproducible, as shown in figure~\ref{fig:XORReproTest}.

As XOR A = NOT A, if we always take the 2$^{\mathrm{nd}}$ point after the first (and only bit in this case) as being the response bit (as we did above for the XOR gate), we have a NOT gate.

\section{Conclusions}

This type of approach is a serial logic gate where the bits are separated in time. This allows us to do logic operations with one memristor at the speed of the spikes (fast) rather than at the speed of equilibration (slow). This approach also allows us to do logic with a two terminal (one-port) device, the extra `complexity' of the operation is contained within the time domain. Essentially, we use the memristor's short-term memory to hold the first bit and do the calculation. This demonstrates that memristors can act as the processor and memory store in one. It also shows the bizarre property of the memory in a system being used to perform memoryless logic. 

The memristor is acting similarly to a sequential logic circuit, where the combinatorial logic is combined with the memory store. Furthermore, the memristor logic gate is asynchronous because there is no need for a clock pulse, but there are issues of race hazard because the second bit must arrive within the time window of the memristor's memory. 

The speed of the these operations is not too fast in this proof-of-principle, however, this is because of the speed at which the electrometer can properly measure a current response. Circuit theory suggests that these spikes should exist at shorter times, so we are confident that the devices can be sped up by sending the second spike in faster. 

The memristor is very low power, especially if operated at the voltages and currents shown in the paper (it is possible to work at higher voltages if desired).

At the moment the output is a different circuit measurable to the input (i.e. the output is current and the input is voltage), it is necessary to convert from one to the other to enable the creation of logical circuits. However, we expect that a current pulse should propagate through a circuit~\cite{MusicPaper} and cause a change in voltage across the next memristor, which could then be used to do the next operation and thus allow the creation of larger memristor logical circuits. We plan to do further work on testing this and investigating the possibility of using a second memristor as a $V \rightarrow I$ transformer (based on the fact that previous $\Delta$V produced a $\Delta$I).

\subsubsection*{Acknowledgments.} This work was funded by EPSRC on grant EP/H014381/1

\end{document}